# SmartHomeSecure: Automated Detection and Repair of Smart Home Configuration Errors Using Large Language Models

Yizhi Wang[1], Xinghua Gao[1*], Reachsak Ly[2], Alireza Shojaei[1]
[1] Myers-Lawson School of Construction, Virginia Polytechnic Institute and State University, Blacksburg, VA 24061, USA
[2] School of Technology, Eastern Illinois University, Charleston, Illinois, United States
[*] Corresponding author, email: xinghua@vt.edu

**Abstract**
Smart home automation platforms increasingly rely on user-authored YAML configuration files to define device behaviors, but these files are prone to syntax, formatting, and semantic logic errors that can cause automation failures and safety risks. Existing YAML validators, static analysis tools, and general-purpose large language models offer limited support for end-to-end diagnosis and repair because they lack domain-specific understanding and validated correction workflows. This paper presents SmartHomeSecure, a prototype for automated detection and repair of Home Assistant configuration errors using lightweight program analysis and constraint-guided large language model generation. SmartHomeSecure parses YAML files, detects syntactic and common semantic errors, normalizes error context, applies deterministic auto-fixes for routine defects, and constructs constrained prompts that guide LLMs toward minimal and structurally valid repairs. The system is implemented as a modular web application with four layers: UI Shell, Feature Orchestrator, Domain Engine, and Integration Layer. Its repair pipeline was evaluated on 100 real-world Home Assistant YAML files with manually injected errors across five categories: syntax/parsing, indentation, mapping, sequence, and scalar quoting errors. Four models were tested: gpt-oss-20b, gpt-oss-120b, llama-3.1-8b, and llama-3.3-70b. Results show that three models achieved 100% error detection accuracy, with repair success rates ranging from 87% to 93%. Manual verification found no hallucinated or incorrect repairs among successful outputs. These findings suggest that combining domain-aware program analysis with constrained generative AI is a feasible approach for improving the reliability and usability of smart home configuration repair.

## 1. Introduction

Smart homes have become a central part of modern living (Schulz & Scilla, 2024). Open-source platforms like Home Assistant, known for local control and high customizability, now serve over two million households worldwide (Home Assistant, 2025). These platforms commonly use YAML files to define complex automation behaviors.

However, the flexibility of YAML configurations also makes them error prone (Anik et al., 2024). Common syntax errors include indentation and formatting issues. Semantic logic errors include invalid triggers and conflicting rules. Such errors can cause automation failures. In worse cases, they lead to abnormal device behavior, which harms user experience and may even compromise safety (Yu et al., 2024).

Despite how common these errors are, no effective tools exist to automatically diagnose and fix them (Anik et al., 2024). Basic syntax checkers cannot understand the domain-specific semantics of smart home configurations, nor can they handle logical errors. Even advanced static analysis tools designed specifically for IoT automation rules have been shown to lack the necessary semantic understanding and common-sense reasoning (Quantrill et al., 2026). A recent study by Anik et al. (2024) examined 190 real-world configuration discussions from the Home Assistant community and found that 68% of the issues involved debugging. Existing tools detected at most 14 out of 129 buggy cases and fixed none.

Users often resort to trial-and-error or community forums for troubleshooting. This process is time-consuming and often ineffective. Even general-purpose large language models do not perform well on this task (Ghaleb & Rathnayake, 2025). Their outputs lack a precise understanding of the configuration context, cannot guarantee correctness, and are prone to hallucination—generating plausible but incorrect code (Lee & Song, 2025). None of these approaches offer an end-to-end automated solution.

To move from error detection to trustworthy repair, this paper proposes SmartHomeSecure, a prototype system for intelligent configuration repair. The system uses generative artificial intelligence as its engine, guided and constrained by domain knowledge. It analyzes an input Home Assistant YAML file, automatically identifies syntax errors and common semantic errors, and works within a constrained environment built from official Home Assistant specifications and best practices. The language model then generates structurally sound and logically coherent repair suggestions, and the system outputs a corrected configuration file. The prototype focuses on the most common and automatable configuration errors.

This paper offers a method, a system, and an empirical evaluation. The method combines lightweight program analysis with constraint guided large language model generation. It extracts the contextual structure of a configuration and injects that structure as precise constraints into prompt engineering, which improves repair accuracy and reduces hallucination in LLM based code repair. The SmartHomeSecure prototype realizes this method in a complete workflow that spans configuration parsing, multi-level error detection, AI guided repair, and validation, demonstrating the engineering feasibility of the approach. The paper also contributes a dataset of Home Assistant configurations with both real and synthetic errors, together with a preliminary evaluation of the prototype's repair capability. The experimental results quantify system performance across common error categories and offer insights into how large language models behave within constrained environments.

The remainder of this paper organizes as follows. Section 2 introduces the technical background and limitations of existing work. Section 3 describes the research methodology and design. Section 4 presents the SmartHomeSecure system framework and its implementation. Section 5 reports experimental design and evaluation results. Section 6 discusses the findings and outlines future work. Section 7 concludes the paper.

## 2. Background

### 2.1 Home Assistant and YAML Configuration Files

Home Assistant is an open-source platform for home automation. Users write its core automation logic in YAML configuration files. YAML is a declarative language that lets users define complex rules involving triggers, conditions, and actions (YAML, 2026). For users without programming experience or professional backgrounds, writing and validating these files manually creates usability and reliability challenges (Zhang et al., 2023). The platform serves as a relevant case study for research on configuration reliability and security in end-user programmable systems.

### 2.2 Types of Configuration Errors

As observed by Anik et al. (2024), configuration defects in Home Assistant automation often involve both syntax violations and semantic logic errors. Syntax errors violate the formal structure of YAML. Examples include incorrect indentation, malformed sequences, and invalid key-value mappings. Semantic logic errors are mistakes in the automation's operational logic. They include invalid API calls, references to non-existent entity states, unreachable actions caused by flawed conditional logic, and conflicts between concurrent automation rules.

### 2.3 Program Analysis and Generative AI

Traditional software program analysis techniques adapt well to YAML-based automation (Yu et al., 2024). The key techniques are control flow analysis and symbolic execution (Cadar & Sen, 2013). Control flow analysis builds a control flow graph to model possible execution paths within an automation rule, typically following the sequence of trigger, condition, and action (Alhanahnah et al., 2020). This modeling helps locate dead code and clarifies the logical dependencies embedded in the configuration. Symbolic execution analyzes automation logic by treating input values as symbolic variables rather than concrete data (Kothari et al., 2008). This method enables systematic exploration of multiple execution paths to uncover runtime errors, edge-case logic flaws, and subtle semantic inconsistencies that may indicate latent vulnerabilities or functional defects. The current prototype does not implement full symbolic execution.

Generative artificial intelligence, particularly large language models (LLMs), offers a way to generate structured code (Pimparkhede et al., 2024). LLMs have demonstrated proficiency in code comprehension, generation, and repair across many programming paradigms (LLaMA, 2026). When provided with appropriate contextual prompts and structural constraints, these models can interpret the intent behind a flawed YAML snippet and generate syntactically correct and semantically reasonable correction candidates. This work uses four LLMs as the core components for generating repair suggestions: LLaMA-3.1-8B, LLaMA-3.3-70B, GPT-OSS-20B, and GPT-OSS-120B. The structured diagnostics produced by the program analysis phase guide and constrain the model outputs.

### 2.4 Limitations of Existing Tools

Current solutions do not fully address the configuration reliability problem. General-purpose YAML validators and online linters typically check only basic syntax. They cannot understand the domain-specific semantics and constraints of the Home Assistant ecosystem (Anik et al., 2024). There are academic research that apply formal methods or advanced static analysis to smart home security, but those approaches often focus on network-level protocols, device firmware vulnerabilities, or communication security (FakhrHosseini et al., 2025; Feng et al., 2024; Kaushik et al., 2025; Nekrasov et al., 2024). They do not directly apply to the end-user-written YAML configuration where logic errors originate.

General-purpose LLM chatbots do not offer a reliable alternative for configuration assistance. Their suggestions are unverified, prone to hallucination (Lee & Song, 2025), and not integrated into a systematic automated validation workflow. A dedicated tool is missing—one that can automatically detect syntactic and semantic errors, propose intelligent corrections grounded in domain knowledge, and validate those corrections for correctness and security. The SmartHomeSecure framework aims to fill this gap.

## 3. Research Methodology and Design

### 3.1 Methodology

The study adopts Design Science Research (DSR) as its research methodology. DSR is a paradigm in information systems research that focuses on solving complex real-world problems by creating and evaluating artifacts (Hevner et al., 2004). Artifacts in this context include constructs, models, methods, frameworks, and instantiated software systems. In this study, the SmartHomeSecure framework and its functional prototype serve as the core artifact.

The core claim of DSR is that new knowledge is generated through the act of designing and evaluating an artifact. This claim distinguishes DSR from research approaches that aim mainly to explain or predict phenomena without building a working solution. The knowledge produced can take the form of design principles, or it can include an understanding of why an artifact works or why it does not work.

This study uses DSR as the sole methodology. This means the entire research revolves around one core question: how to build a system that can automatically repair smart home configuration errors. All technical decisions, including lightweight program analysis and constraint-guided large language model generation, were made within the DSR framework.

A central tenet of DSR is that it generates new knowledge through the act of designing and evaluating an artifact. This distinguishes DSR from research approaches that aim primarily to explain or predict phenomena without building a working solution. The artifact itself becomes the vehicle for knowledge contribution. The knowledge produced can take the form of design principles, an understanding of what makes the artifact effective, or lessons about what does not work and why. The study also seeks to generate design knowledge beyond analytical findings.

The research process follows the Design Science Research cycle, which includes the iterative phases of problem identification, definition of solution objectives, design and development, demonstration, evaluation, and communication (Hevner et al., 2004). These phases are not strictly sequential. Researchers often return to earlier phases as new insights emerge during later stages. The iterative nature of DSR allows the artifact to be progressively refined based on empirical feedback. Each cycle improves the fit between the artifact and the problem it is meant to solve.

In this study, the DSR cycle is executed as follows. The problem identification phase used literature review and community discussions to determine that Home Assistant YAML configuration errors lack automated repair tools. The objective definition phase specified the necessary capabilities of a solution: detecting syntax and semantic errors, generating context-aware repair suggestions, and validating corrected configurations. The design and development phase completed the layered architecture design of SmartHomeSecure and implemented the prototype, including the Domain Engine Layer and the Integration Layer. The demonstration phase tested the prototype on a dataset of 100 configuration files with injected errors. The evaluation phase measured error detection accuracy and repair success rates across four large language models and manually verified all results. The communication phase reports the design knowledge and research findings through this paper.

Gregor & Hevner (2013) propose a framework for positioning design science contributions based on problem and solution maturity. When a problem is well recognized but existing solutions remain inadequate, the research falls into what they term the improvement quadrant. The field of smart home configuration repair occupies this quadrant. The prevalence and impact of YAML configuration errors are well documented in both academic literature and community discourse, yet accessible automated tools for diagnosis and repair do not exist. SmartHomeSecure enters this space as an improvement-oriented artifact. The study does not set out to identify a new problem. It aims to construct a solution where current tools fall short. This objective of designing and building an innovative intervention for a recognized need represents a typical context for Design Science Research (Hevner et al., 2004).

DSR functions as a methodological framework rather than a technical prescription. It does not dictate the specific algorithms or technologies that must be used inside the artifact. What DSR requires is that the work originates from a clearly identified practical need, that a working system is produced, and that the system is evaluated in a way that yields credible evidence about its effectiveness. In the present study, the technical methods integrated into SmartHomeSecure—lightweight program analysis and constraint-guided large language model generation—were selected and refined within this DSR cycle. The design and evaluation of these methods are guided by the DSR requirement that an artifact must demonstrate practical utility while also yielding insights that can inform future work. DSR provides the overarching logic that connects the technical choices to the study's broader contributions.

Design Science Research provides a formal structure for this iterative process, keeping the work focused on utility and effectiveness (Seymour et al., 2026). Another important reason for using DSR in this study is that DSR allows researchers to build a functional software system as the main research output within an academic framework. The problem of smart home configuration repair requires an actual working system to verify the feasibility of the proposed method. Theoretical analysis or pure data observation alone cannot prove the effectiveness of automatic repair. DSR legitimizes this engineering-intensive research approach and provides criteria for evaluating such artifacts, including utility, reliability, and demonstrability.

### 3.2 Research Design

The research process organizes into three connected phases: problem identification and objective definition, framework design and prototype development, and demonstration with experimental evaluation. This three-phase structure follows established design science research process models, such as the six-step methodology proposed by Peffers et al. (2007) and Hevner et al. (2004). Recent discussions of design science methodologies note that different research contexts may require flexible entry points and adaptive procedures (Seymour et al., 2026).

Design Science Research (DSR) is chosen as the methodology for this research design because the problem is already well documented but lacks existing solutions. The three phases directly implement the DSR cycle of problem identification, artifact creation, and evaluation. This structure ensures that every research activity serves the goal of building and testing a working artifact, rather than only describing or explaining the problem.

The problem identification draws on analysis of industry reports, academic literature, and user community discourse. Errors in end-user smart home configurations are common and impactful, but accessible automated tools for diagnosis and repair are lacking. This situation matches the DSR "improvement quadrant" defined by Gregor & Hevner (2013). The problem is real and relevant, yet no mature solution exists. DSR is therefore appropriate because it guides the researcher to construct a new artifact instead of only analyzing existing failures. The primary design objective is a software that can automatically detect and correct both syntactic and common semantic errors within Home Assistant YAML configurations. This objective requires several core capabilities: accurate error detection, generation of context-aware repair suggestions, and a mechanism for preliminary validation of corrected configurations. Within DSR, defining such measurable objectives is a required step before design begins. This step keeps the artifact tied to practical needs and prevents the research from drifting into purely technical exploration.

The framework then undergoes conceptual design, and a functional prototype emerges. The solution integrates lightweight program analysis with generative artificial intelligence. In this design, program analysis provides structural and semantic context and offers verifiable constraints, while the generative AI component produces syntactically valid repair candidates within those constraints. This conceptual design translates into a working prototype that implements the core SmartHomeSecure pipeline. The implementation includes a dedicated configuration parser, a hybrid error detection module that uses both rule-based checks and lightweight static analysis, a repair generation module that employs constraint-guided prompt engineering to interface with the language model, and a validation component that assesses the structural integrity of generated repairs. DSR supports iterative design. Early versions of the prototype used only LLM repair without program analysis, which produced hallucinated fixes. After evaluating those failures, the design was refined by adding the Domain Engine Layer. The core mechanism of Design Science Research (DSR) is the cycle of design, demonstration, evaluation, and redesign. DSR does not treat iteration as a failure but as a way to improve the artifact's fit to the problem. The final design combines deterministic auto-fix with constrained LLM generation have emerged from this iterative process.

The prototype demonstrates its end-to-end capabilities by running on a curated dataset of representative Home Assistant configurations containing known flaw types. After this demonstration, an experimental evaluation assesses the artifact's effectiveness. Section 5 presents the detailed methodology and results. The evaluation metrics center on quantitative performance indicators, mainly error detection accuracy and repair success rate. The outcomes of this evaluation serve two purposes: verifying how well the initial design objectives are met and providing empirical evidence for future iterations and refinements of the framework. DSR requires that an artifact be evaluated with credible evidence. The evaluation design in this study uses a dataset of 100 files, four large language models, and manual verification of every repair. This design follows DSR guidelines. The evaluation not only tests the prototype but also generates design knowledge. For example, the results show that exposing structural constraints to the LLM improves repair accuracy and that larger models perform better but even smaller models can work well when constraints are provided. This knowledge can inform future research on AI-assisted configuration repair.

The research design as a whole explains why DSR is used. The problem is practical and well recognized. The solution is a working artifact built through iteration. The evaluation is rigorous and empirical. And the outputs include both a functional system and generalizable design knowledge. These characteristics are what DSR is designed to produce. The three-phase research design is not just a convenient structure. It is a direct implementation of the DSR methodology.

## 4. System Framework

The source code of the first prototype of SmartHomeSecure has been released on GitHub at https://github.com/reachsak/SmartHomeSecureProject. As the first working artifact produced within the Design Science Research cycle, this prototype serves as the foundational codebase for ongoing and future research.

### 4.1 Overall Architecture

The SmartHomeSecure framework adopts a layered architecture. This separation of concerns enhances modularity and maintainability. Figure 1 illustrates the overall architecture. The system comprises four primary layers. The UI Shell Layer handles user interaction, routing, and visual feedback. The Feature Orchestrator Layer manages model selection, file upload workflow, and the repair state machine. The Domain Engine Layer encapsulates core YAML validation, error diagnosis, and deterministic auto-fix logic. The Integration Layer provides a unified interface to external LLM providers and handles API communication. This architectural decomposition directly implements the constraint-guided LLM generation method proposed in the introduction. The Domain Engine Layer provides precise structural and semantic constraints extracted from the configuration context. Guided by these constraints through prompt engineering, the Integration Layer interfaces with the generative capabilities of the LLaMA models to produce contextually appropriate repairs.

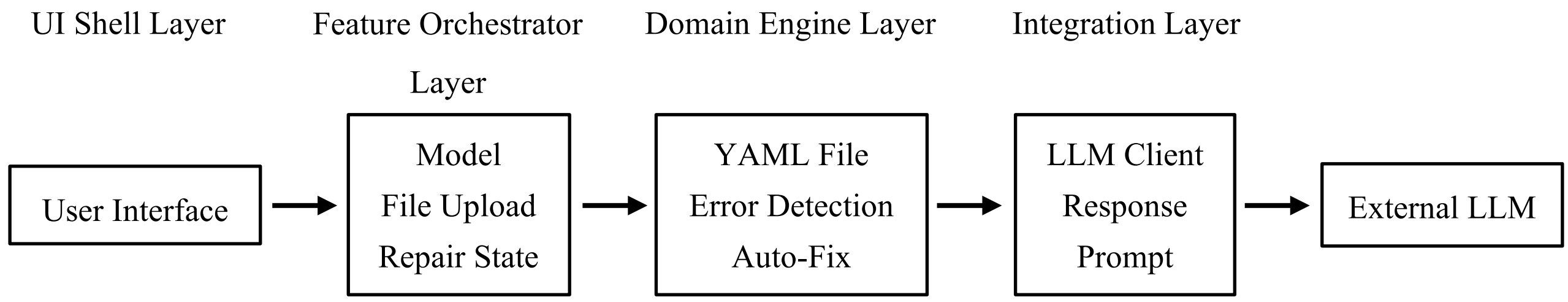

Figure 1. Overall architecture of SmartHomeSecure

### 4.2 UI Shell Layer

The UI Shell establishes the foundational application structure and navigation paradigm. This layer resides within the main application component. It owns the routing logic that maps URL paths to specific views. The primary AI assistant view remains accessible through both the root path and an explicit route. The header component provides consistent branding and navigation links, maintaining a uniform user experience across the application. This layer intentionally maintains minimal state. It delegates feature-specific responsibilities to the underlying orchestrator layer, adhering to the principle of separation of concerns. Figure 2 shows the main user interface of the SmartHomeSecure prototype. The interface provides access to model selection through a tabbed layout.

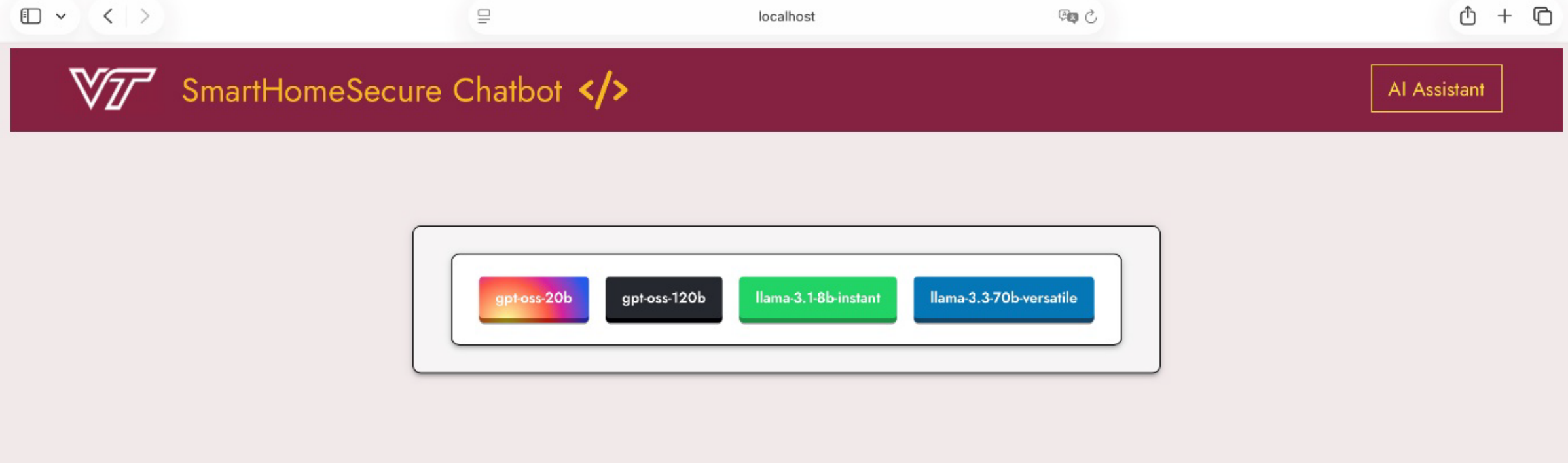

Figure 2. SmartHomeSecure user interface

### 4.3 Feature Orchestrator Layer

The Feature Orchestrator layer serves as the system's control center. It manages the end-to-end workflow from model selection to corrected file export. The main orchestrator component acts as the primary model switchboard. It renders a tabbed interface that allows users to select among four available Large Language models. Each module maintains its own prompt history state, file selection state, and conversation context. This design enables independent operation while conforming to a shared interface contract.

The orchestrator layer implements a state machine for the upload file and repair workflow. After file selection, the component invokes validation routines from the Domain Engine Layer. It tracks the number of repair attempts against a configurable maximum. It manages the progression from initial validation through automatic fixes to AI-mediated repair and output generation. This state management ensures that users receive clear feedback on the status of each uploaded file, whether successful validation, auto-fix application, AI-guided correction, or unrecoverable failure. The orchestrator also coordinates the download of corrected YAML files and the generation of optional fix reports, providing transparency into the performed modifications.

### 4.4 Domain Engine Layer

The Domain Engine Layer forms the intellectual core of the SmartHomeSecure framework. It contains all domain-specific knowledge about Home Assistant YAML configurations and implements the program analysis techniques discussed in Section 2.3. This functionality resides in a dedicated utility module. The layer provides a comprehensive suite of functions for YAML processing, error diagnosis, and deterministic repair.

The validation subsystem starts with a custom YAML parser built upon a standard YAML library. The parser extends the standard library to accommodate Home Assistant's specific schema requirements and custom tags. The main validation function performs syntactic validation against the YAML specification. At the same time, it checks for domain-specific correctness, such as proper trigger-action structure and valid entity references. When validation fails, the error normalization component translates parser exceptions and validation failures into a standardized error format. This format captures error type, location, and context. This normalized representation enables consistent handling across different error categories and provides the foundation for the constraint extraction mechanism.

Building upon the normalized error representation, the context extraction subsystem analyzes the surrounding YAML structure to recover the intended configuration semantics. This component implements lightweight program analysis techniques adapted from traditional software analysis. One technique constructs a simplified control flow graph for automation rules to identify the sequence of trigger, condition, and action elements. The structural hint generation component then produces machine-readable specifications of the expected configuration patterns. These hints encode domain constraints that guide subsequent LLM interactions. They include information about required fields, valid value ranges, entity type compatibility, and typical automation patterns drawn from Home Assistant best practices.

The deterministic auto-fix primitives constitute the final component of the Domain Engine Layer. These routines address common, well-understood error patterns through rule-based transformations. They do not involve the LLM. Examples include correcting indentation inconsistencies, repairing common key misspellings, and restructuring malformed trigger definitions. The auto-fix function iteratively applies these primitives, revalidating after each transformation and accumulating a record of applied fixes. This deterministic layer resolves many simple errors efficiently and reduces the burden on the LLM, allowing the model to focus on more complex semantic repairs that require genuine understanding and generation capability.

**4.5 Integration Layer**

The Integration Layer abstracts communication with external LLM providers. The system configures it with appropriate parameters including model selection and maximum token limits. The current implementation operates in a client-side capacity. This choice facilitates rapid prototyping and demonstration while acknowledging the security considerations discussed in Section 6.

The prompt engineering subsystem within this layer translates the structured constraints produced by the Domain Engine into natural language instructions for the LLM. The repair prompt follows a careful template. It includes the original flawed YAML snippet, the normalized error description with location information, structural hints about the expected configuration pattern, and explicit instructions for minimal edits. This approach embodies the constraint-guided generation principle. It embeds the program analysis results directly into the conversational context, effectively constraining the LLM's generative space to plausible repairs. The prompt explicitly instructs the model to preserve all functionally correct portions of the configuration while addressing only the identified issues. This instruction minimizes the risk of introducing new errors during the repair process.

The response parsing component handles the variability inherent in LLM outputs. A multi-stage extraction strategy handles this variability. The system first attempts to parse the complete response as valid YAML. If this fails, it employs pattern matching to extract YAML code blocks from markdown-formatted responses. Fallback mechanisms progressively relax parsing assumptions while maintaining structural integrity checks. This design ensures that the system can recover useful repairs even from imperfectly formatted model outputs. The extracted YAML then undergoes revalidation through the Domain Engine Layer, creating a feedback loop that enables iterative refinement when initial repair attempts prove insufficient.

**4.6 Workflow Logic**

Figure 3 depicts the complete repair workflow. As implemented in the model-specific chat components, the workflow follows a systematic progression from file ingestion to corrected output. After a user uploads a file, the system starts the Domain Engine's validation routines. If the YAML validates no error exist, the workflow terminates early and notifies the user of the configuration's correctness. For invalid configurations, the system first engages the deterministic auto-fix subsystem. It applies rule-based transformations and revalidates after each fix. This phase continues until either validation succeeds or no further automatic fixes can be applied.

When deterministic fixes prove insufficient, the system constructs a comprehensive repair request for the LLM. The request incorporates the original YAML, the complete error context extracted during validation, and the structural hints generated by the Domain Engine. The Integration Layer transmits this request to the selected LLM model, receives the response, and passes the extracted YAML back to the Domain Engine for revalidation. If validation fails, the system increments the attempt counter. If the maximum attempts have not been exceeded, it constructs a new prompt that includes the previous attempt's output and the remaining errors. This iterative refinement continues. After successful validation or exhaustion of attempts, the system presents the outcomes to the user. It distinguishes between originally valid, auto-fixed, AI-repaired, and failed configurations. Users may then download individual corrected files or a comprehensive fix report documenting all modifications. Figure 4 shows an example fix report generated by the system. The report lists each detected error, its location, the applied fix, and the validation status after repair.

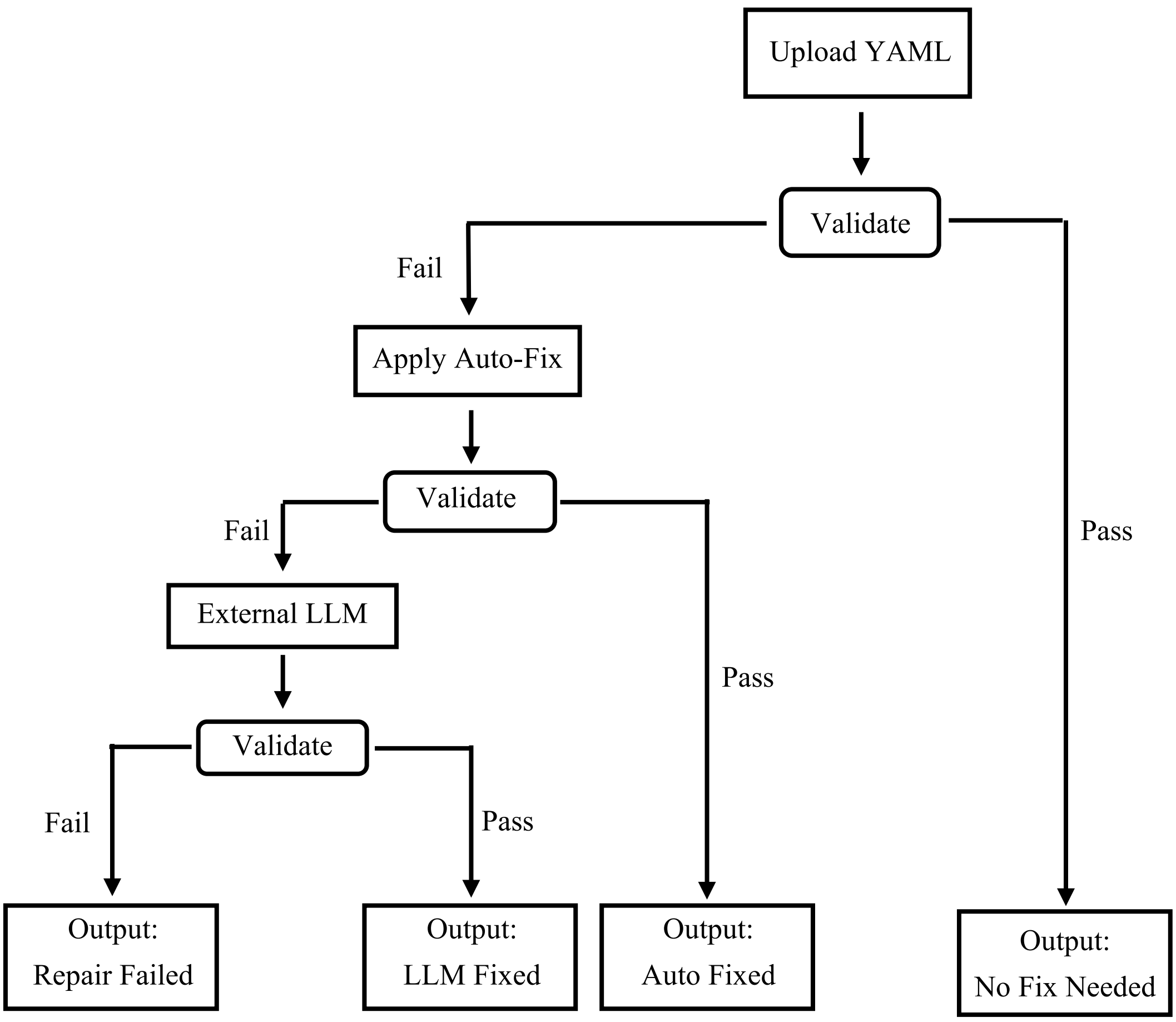

Figure 3. SmartHomeSecure repair workflow

fix-report-131.txt

```
---
File: bedroom_off.yaml
Source: auto
Attempts: 0
Issues: 1
Cause: tab characters must not be used in indentation (line 9, column 1)
Action: Auto-fix corrected the YAML syntax
Status: Fixed
Notes: Converted leading tabs to spaces for YAML indentation.
---
File: binary_sensors.yaml
Source: auto
Attempts: 0
Issues: 1
Cause: tab characters must not be used in indentation (line 18, column 1)
Action: Auto-fix corrected the YAML syntax
Status: Fixed
Notes: Converted leading tabs to spaces for YAML indentation.
---
File: blink.yaml
Source: auto
Attempts: 0
Issues: 1
Cause: tab characters must not be used in indentation (line 10, column 1)
Action: Auto-fix corrected the YAML syntax
Status: Fixed
Notes: Converted leading tabs to spaces for YAML indentation.
---
```

Figure 4. Example fix report

## 5. Evaluation

### 5.1 Evaluation Goal

The experimental evaluation aims to measure how effectively the SmartHomeSecure prototype detects and repairs common syntax and semantic errors in Home Assistant YAML configurations. Detection accuracy refers to the framework's ability to correctly identify the presence and type of an injected error in a given configuration file, which depends on the rule-based syntax checking and lightweight static analysis implemented in the Domain Engine Layer. Repair accuracy refers to the framework's ability to generate a corrected configuration that is both syntactically valid and semantically faithful to the original automation logic. The evaluation also examines how four different large language models perform under the same constraint-guided repair pipeline, and whether the system ever produces incorrect or hallucinated repairs. The goal is not to compare against other tools—since no comparable end-to-end repair tool exists—but to provide a controlled, reproducible assessment of the framework's core capabilities and to generate empirical evidence for the feasibility of the proposed approach.

### 5.2 Experimental Design

A multi-stage experimental procedure addresses these questions. The key steps include dataset creation, error injection, framework execution under multiple model configurations, and manual verification of results.

The study curates 100 YAML files from a public community repository on GitHub named HA_Blueprints, maintained by SirGoodenough. The repository is publicly available at https://github.com/SirGoodenough/HA_Blueprints. The repository contains real-world automation and script blueprints, including examples for device control, notification delivery, and complex conditional logic. File sizes range from approximately 100 to over 3,000 lines. This range represents the complexity and diversity found in user-authored configurations.

To establish a controlled baseline for measuring detection and repair capabilities, the experiment deliberately introduces a single predefined error into each of the 100 files. These five error types are defined based on the taxonomy in Section 2.2. Syntax/parsing errors include misspelled YAML keywords, missing commas, or improper use of colons that break the basic structure. Indentation and whitespace errors occur when spaces or tabs are misaligned, violating YAML's block scoping rules. Mapping errors refer to malformed key-value pairs, such as missing colons, duplicate keys, or incorrect nesting of mappings. Sequence errors involve improper list formatting, for instance missing dash prefixes or mixing sequence and mapping syntax incorrectly. Scalar quoting errors arise when strings require quotes but lack them, or when quotes are used inconsistently.

Each error was injected manually into a clean, validated copy of the original configuration file. A precise mutation rule was defined for each error type to ensure consistent injection across all 100 files. For example, a syntax error might delete a required colon, while an indentation error shifts a block by two spaces. After injection, every file was independently inspected to confirm that exactly one error was present and the rest of the configuration remained syntactically correct and semantically identical to the original. This rigorous injection and verification process allows the experiment to measure detection and repair accuracy without confounding factors such as multiple interacting errors or pre-existing defects. For each file, the precise location and type of the error are known, and the corresponding correct version is available for comparison.

The entire experiment runs on a MacBook Pro with an Apple M4 chip and 16GB RAM. The SmartHomeSecure prototype, implemented as a React-based web application as described in Section 4, processes all test files. To explore the impact of the generative AI engine, the experiment repeats the process using four different large language models: a 20-billion-parameter general-purpose model (gpt-oss-20b), a larger 120-billion-parameter model (gpt-oss-120b), and two of Meta's LLaMA models (llama-3.1-8b and llama-3.3-70b). For each model, the system processes all 100 error-injected files. The system first parses each file and the error detection module. Upon detecting an error, it constructs a prompt incorporating error context information, employing the constraint-guided prompt engineering method described in Section 4. For overly long files exceeding token limits, the system isolates the error context by extracting the erroneous line along with 100 lines of code above and below it before sending the context to the LLM for repair generation. The system then reintegrates the repaired code.

After the repairs are complete, we manually verify each repaired file against the original correct version to confirm the accuracy of the fix and check both syntactic validity and semantic equivalence. Semantic equivalence means the automation logic remains functionally identical to the original intended behavior. In practical terms, this requires that the repaired configuration preserves the exact sequence of triggers, conditions, and actions from the original valid version. For example, if the original automation was meant to turn on a light when motion is detected and the sun is below the horizon, the repaired version must still contain the same motion trigger, the same sun condition, and the same light action in the same logical order. Any repair that changes the trigger-condition-action flow or adds unintended side effects is considered semantically incorrect and therefore unsuccessful. Only repairs that meet both criteria count as successful.

### 5.3 Results and Analysis

Table 1 summarizes the performance results of the SmartHomeSecure framework using the four LLMs.Regarding error detection accuracy, aside from the gpt-oss-20b model, the other three models demonstrate very high effectiveness. The gpt-oss-120b, llama-3.1-8b, and llama-3.3-70b models successfully detect the injected errors in all 100 test files, achieving 100 percent detection accuracy. This success comes from the framework's robust pre-detection module, which combines rule-based syntax checking with lightweight static analysis to reliably identify defects.

The gpt-oss-20b model performs poorly at this stage. It incorrectly reports zero errors in 12 samples. It is important to note that the framework's built-in error detection module in the Domain Engine Layer successfully identified the injected errors in all 12 samples before sending the error context to the LLM. The failure occurred after the detection phase. When the error information was passed to the gpt-oss-20b model for repair generation, the LLM either returned an empty response or produced output that could not be parsed as YAML, reporting that no errors were found. Further analysis of the logs reveals that the 12 failing samples are not concentrated in a single error category. The injected error types are distributed across all five categories, with syntax errors and mapping errors appearing slightly more frequently. However, all 12 samples share a common characteristic. They are among the longest files in the dataset, each exceeding 2,000 lines. For such long configurations, the gpt-oss-20b model often fails to extract the minimal error context correctly, either exceeding its effective context window or generating malformed responses that omit the repaired YAML block altogether. By contrast, the larger models handled these lengthy files without issue. This result indicates that the 20-billion-parameter model lacks sufficient capability to understand the configuration context and generate meaningful repairs, even when the detection module has already pinpointed the error. This outlier case highlights the importance of having a stable, model-agnostic pre-detection layer and selecting LLMs with adequate capacity for code repair tasks, especially when processing large configuration files.

Regarding repair success rates, the results are similarly positive. For the three models that achieve 100% error detection rate, the repair success rates are 93 percent for gpt-oss-120b, 87 percent for llama-3.1-8b, and 91 percent for llama-3.3-70b. Manual verification confirms that all generated repairs are accurate. No instance occurs of the LLM introducing new errors or hallucinating incorrect fixes. The exact alignment between the number of reported fixes and the number of manually verified fixes demonstrates that whenever the framework generates a repaired file, it is consistently correct. The manual verification found no instances of hallucinated fixes in the tested cases.

The variation in success rates among different models reflects inherent differences in code understanding capabilities. The 120-billion-parameter model outperforms the smaller 8-billion-parameter LLaMA model. Larger models generally perform better. The entire experiment runs smoothly on a MacBook Pro with an Apple M4 chip and 16GB of memory. This result indicates that the framework has good practicality and deploy ability.

**Table 1. Performance of different LLMs on error detection and repair**

| LLM model | Error detection accuracy | Repair success rate | Successful repairs |
| --- | --- | --- | --- |
| gpt-oss-20b | 88% (12 files reported zero error) | 85 % | 85 |
| gpt-oss-120b | 100% | 93 % | 93 |
| llama-3.1-8b | 100% | 87 % | 87 |
| llama-3.3-70b | 100% | 91 % | 91 |

*Note: Error detection accuracy for gpt-oss-20b is 88% because the model reported zero errors in 12 files where errors actually existed; for the other three models, all errors were correctly detected. Repair success rate refers to the proportion of files for which the model generated a correct, verified fix.*

## 6. Discussion

### 6.1 Interpreting the Experimental Outcomes

The experimental results show clear patterns. With three of the four language models, SmartHomeSecure detects errors with 100 percent accuracy. The repair success rates for those three models range from 87 to 93 percent. The system produces no incorrect repairs. These results help explain why the approach works and what the variation across models suggests about the configuration repair task.

The alignment between the number of reported fixes and the number of manually verified fixes across all successful models is a notable finding. It indicates that the constraint-guided generation method effectively reduces hallucination in LLM-based code repair.

The results also indicate a role for the pre-detection layer. Three models achieve high detection rates while the fourth (gpt-oss-20b) fails on 12 files. This observation suggests that the framework's reliability depends on a stable, model-agnostic frontend that can consistently extract and normalize error information. When this frontend functions correctly, the LLM's task becomes more tractable.

The variation in repair success rates across models shows a non-linear relationship between model scale and repair capability. The 8-billion-parameter LLaMA model achieves an 87 percent success rate, six percentage points below its 70-billion-parameter counterpart. This relatively small gap suggests that the constraint-guided approach reduces the raw generative capacity required for successful repair. When the problem is well-structured and the context is rich, even smaller models can perform competently.

The failure of gpt-oss-20b warrants attention. The model incorrectly reports zero errors in 12 files. This pattern is consistent with either a failure in the detection module's interaction with this specific model or a limitation in the model's ability to parse and respond to structured repair prompts. Given that the other three models succeed under identical conditions, a plausible explanation lies in the model's training distribution or instruction-following capability. This result shows that not all LLMs are equally suited for code repair tasks, and model selection remains an important design decision in systems of this kind.

### 6.2 Theoretical Implications

Beyond the empirical findings, this study contributes to an understanding of AI-assisted configuration repair in end-user programmable systems. The effectiveness of LLM-based repair increases when the system exposes, rather than hides, the structural and semantic constraints of the target domain. Implicit knowledge embedded in the LLM's training data does not suffice. The prompt must contain explicit, verifiable constraints.

A repair system may benefit from a three-tier fallback hierarchy: deterministic rules for routine errors, constrained LLM generation for complex but well-understood errors, and human-in-the-loop for ambiguous or new cases. The current prototype implements the first two tiers. The third remains for future work.

### 6.3 Limitations and Critical Reflection

The evaluation covered a limited range of error types. The evaluation focused on a predefined set of five common error types. While these represent a portion of real-world configuration issues, they do not cover the full spectrum of possible flaws. Errors involving cross-automation interference, temporal logic flaws such as timing-dependent conditions, and security policy violations remain outside the current scope. The framework's ability to handle these more complex error categories does not allow inference from the current results.

The test dataset also has limited representativeness. The test dataset, while diverse in file size and structure, came from a single public repository. Configurations authored by the broader Home Assistant community may exhibit greater variability in style, complexity, and error patterns. The deliberate injection of single errors per file, while methodologically necessary for establishing ground truth, does not reflect the reality of configurations that may contain multiple interacting errors. The framework's performance under such conditions remains untested.

The depth of validation is another limitation. The current validation mechanism verifies syntactic correctness and basic structural integrity, but it does not execute the repaired configuration in a simulated or actual Home Assistant environment. Errors that only manifest at runtime, such as race conditions, device unavailability, or unexpected state transitions, remain outside the framework's detection capability. The absence of dynamic validation means that repaired configurations may still fail when deployed.

There are limitations regarding the security and deployment considerations. The client-side architecture, with exposed API keys and no backend isolation, makes the current prototype unsuitable for production use. Beyond the credential security issue, this design also precludes rate limiting, audit logging, and multi-user isolation. A production-ready system would require complete architectural refactoring.

### 6.4 Application Scenarios

The SmartHomeSecure framework demonstrates practical utility across several distinct stakeholder groups. Below, the value for end users, developers, and educators and researchers is elaborate.

For end users, the system offers an accessible way to diagnose and repair configuration errors without requiring deep technical expertise. Many smart home users are not professional programmers. They rely on community forums, trial-and-error, or external support (Anik et al., 2024). SmartHomeSecure lowers the barrier to entry by providing an automated tool that can integrate into existing smart home dashboards or operate as a standalone web service. Users can upload a problematic YAML file and receive a corrected version, along with an explanation of what was fixed. This can save time and reduce the risk of insecure or malfunctioning automations that could compromise home safety, for example, a misconfigured door lock automation that fails to engage at night. Over time, wider use of such tools could reduce the number of configurations-related support requests in smart home communities.

For developers, the framework can integrate into Home Assistant or similar platforms as a diagnostic and repair module. Platform developers currently face a support burden due to user-submitted configuration errors, which often stem from subtle YAML syntax or logic mistakes. By embedding SmartHomeSecure as a built-in tool, developers can offer users real-time feedback and one-click repairs directly within the platform's user interface. The constraint-guided repair approach can adapt to other configuration languages used in the ecosystem, such as those for add-ons or custom integrations. This can reduce the need for manual debugging by developers, free up community support channels, and improve user retention. A platform that offers intelligent self-repair capabilities may gain an advantage in usability and reliability.

For educators and researchers, the prototype serves as a example of a broader class of AI-assisted programming tools for domain-specific languages. Educators teaching courses on software engineering, human-computer interaction, or IoT development can use SmartHomeSecure as a case study to illustrate how program analysis and generative AI can combine to solve real-world problems. Students can explore the codebase, modify the constraint extraction rules, or experiment with different LLMs to understand tradeoffs in repair accuracy and efficiency. For researchers, the framework provides a platform for further experimentation, for example, investigating the effectiveness of alternative prompt engineering strategies, studying the types of configuration errors that remain challenging for LLMs, or conducting comparative evaluations of different model architectures. SmartHomeSecure provides a methodological blueprint that can transfer to other end-user programming domains, such as industrial automation, cloud infrastructure configuration, or business process modeling.

**6.5 Future Research Directions**

The findings and limitations of this study point toward several avenues for future investigation. Future work should extend error coverage to include multifile and temporal logic flaws. The current framework handles single-file syntax and common semantic errors, but real-world smart home configurations often involve multiple interdependent files, such as separate YAML files for automations, scripts, and templates. Errors can propagate across these files, and a change in one may break functionality in another. Temporal logic flaws, for example, automation that assumes a certain order of events without proper synchronization, are common but currently not detected. To address this, future work can extend the program analysis module to construct a cross-file control flow and data dependency graph. Symbolic execution techniques can then apply to reason about temporal properties, such as ensuring that one action always happens before another. The impact of this extension would be to move the framework from isolated file repair to holistic system repair, enabling detection of race conditions, deadlocks, and causality violations. This would bring smart home configuration repair closer to the level of safety-critical system verification.

Integrating simulation-based runtime validation would address the current lack of dynamic checking. Static validation, as implemented in the current prototype, cannot catch errors that only manifest during execution, such as device unavailability, unexpected state transitions, or resource exhaustion. Integrating a simulation environment, for example, a virtual Home Assistant instance with mock devices, would allow the framework to execute repaired configurations in a controlled sandbox before presenting them to the user. The simulation would run a set of predefined test scenarios, such as motion detected at night or temperature exceeding a threshold, and observe whether the automation behaves as intended. Errors detected during simulation can trigger another repair iteration. The methodology involves building an event simulator that mimics sensor inputs and device states, coupled with a monitoring system that checks expected outcomes. Simulation-based validation could increase the reliability of repairs, reducing the risk of deploying a fixed configuration that still fails in practice. It would also enable regression testing across multiple configurations, helping to ensure that fixing one error does not introduce another.

Introducing a human-in-the-loop repair mode would handle ambiguous or low-confidence cases. Not all configuration errors are well-understood or uniquely fixable. In some cases, multiple repairs are semantically plausible, and the LLM's confidence in any single repair is low. The current framework simply fails after exhausting attempts. A more robust approach is to introduce a human-in-the-loop mode, where the system presents the user with diagnostic information and a ranked list of candidate repairs, allowing the user to select or refine the correct one. The methodology includes confidence estimation from the LLM using token probabilities or ensemble sampling, a user interface that visualizes the differences between original and repaired code, and a feedback mechanism that records user decisions to finetune the model for similar cases in the future. It extends the applicability of the framework to errors that are currently out of scope, and it creates a learning loop where user corrections can improve the underlying model, potentially reducing the need for human intervention over time.

Cross-domain generalization to other configuration languages would demonstrate the broader applicability of the constraint-guided generation method. The method is not inherently tied to Home Assistant or YAML. There are many other domains that rely on declarative configuration languages that suffer from similar errors. Errors can lead to service outages, security breaches, or cost overruns. Future work can adapt the SmartHomeSecure framework to these domains by replacing the domain-specific parser and constraint extractor while keeping the core repair pipeline intact. The methodology involves building a library of domain-specific schemas and error patterns, then training or fine-tuning the LLM on each domain's configuration corpus. A generalizable configuration repair tool could deploy across cloud computing, network management, and IoT, potentially reducing human error in infrastructure as code. This would position the framework as a technology for reliable automation in the broader software and systems engineering landscape.

## 7. Conclusion

This paper addressed the problem of configuration errors in smart home automation systems, specifically focusing on Home Assistant YAML configurations. Many smart home users lack effective tools to automatically detect and repair syntax and semantic errors in their automation rules. Existing solutions, including general-purpose YAML validators and unguided large language model interactions, do not provide reliable, context-aware, end-to-end repair capabilities.

To address this gap, this paper proposed SmartHomeSecure, a framework that integrates lightweight program analysis with constraint-guided large language model generation. The framework first validates YAML configurations, extracts structural and semantic context, and applies deterministic auto-fixes for routine errors. When deeper repair is needed, it constructs a constrained prompt that embeds error location, type, and structural hints. It then invokes an LLM to generate a corrected configuration. The prototype system exists as a modular web application, demonstrating the engineering feasibility of this approach.

This research makes several contributions. A constraint-guided LLM generation method improves repair accuracy and reduces hallucination by injecting precise program analysis results into prompt engineering. The SmartHomeSecure prototype realizes a complete workflow from configuration parsing and error detection to AI-guided repair and validation. A dataset of 100 YAML files with injected errors provides a basis for evaluation. The preliminary evaluation using four different LLMs provides empirical evidence of the framework's effectiveness.

The experimental results show the following. With three of the four language models, SmartHomeSecure detects errors with 100 percent accuracy. The repair success rates for those three models range from 87 to 93 percent. The system produces no incorrect repairs. Manual verification confirms that all reported fixes are syntactically valid and semantically faithful to the original automation logic. These findings indicate that combining program analysis with constraint-guided LLM generation offers a feasible approach for automating smart home configuration repair.

The current prototype has limitations. The evaluation covers only five common error types. The dataset comes from a single repository. No runtime simulation validates dynamic behavior. The client-side architecture requires refactoring for production security. Future work will extend error coverage to multi-file and temporal logic flaws, integrate simulation-based validation, explore human-in-the-loop repair, and test cross-domain generalization to other configuration languages.

In summary, SmartHomeSecure shows that automated repair of smart home configurations can succeed by combining program analysis with constrained generative AI. This work provides a foundation for building more robust and user-friendly tools to enhance the reliability of end-user programmable smart home systems.